\documentclass{aastex63}

\usepackage{graphicx,epsfig}% Include figure files
\usepackage{dcolumn}% Align table columns on decimal point
\usepackage{bm}% bold math
\usepackage{xcolor}
\usepackage{amsmath}
\newcommand{\be}{\begin{equation}}
\newcommand{\ee}{\end{equation}}
\newcommand{\bse}{\begin{subequations}}
\newcommand{\ese}{\end{subequations}}
\newcommand{\bary}{\begin{eqnarray}}
\newcommand{\eary}{\end{eqnarray}}
\newcommand{\bwt}{\begin{widetext}}
\newcommand{\ewt}{\end{widetext}}
\begin{document}

\title{Photohadronic model for the neutrino and $\gamma$-ray emission from TXS 0506+056}

\correspondingauthor{Carlos E. L\'opez Fort\'in}

\author{Sarira Sahu}
\email{sarira@nucleares.unam.mx}

\author{Carlos E. L\'opez Fort\'in}
\email{carlos.fortin@correo.nucleares.unam.mx}
\affiliation{Instituto de Ciencias Nucleares, Universidad Nacional Aut\'onoma de M\'exico, \\
Circuito Exterior, C.U., A. Postal 70-543, 04510 Mexico DF, Mexico}

\author{Shigehiro Nagataki}
\email{shigehiro.nagataki@riken.jp}
\affiliation{Astrophysical Big Bang Laboratory, RIKEN,\\
Hirosawa, Wako, Saitama 351-0198, Japan}
\affiliation{Interdisciplinary Theoretical \& Mathematical Science (iTHEMS),\\
RIKEN, Hirosawa, Wako, Saitama 351-0198, Japan}

\begin{abstract}
The detection of a high energy muon neutrino on 22 September 2017 by
IceCube neutrino detector coincides with the multiwavelength flaring
from the BL Lac object TXS 0506+056, most likely confirming AGN as a source of high
energy cosmic rays and neutrinos. Using the photohadronic model, we have explained the very high energy $\gamma$-rays 
observed by MAGIC telescopes few days after the neutrino event and extend the model to calculate the neutrino flux at different windows consistent with the flaring period of TXS 0506+056 and compared with the IceCube and MAGIC estimates. We also use this model to estimate the neutrino flux from the flaring of FSRQ PKS B1424-418 which is believed to be associated with the 2 PeV neutrino event observed by IceCube.
\end{abstract}


\section{Introduction} \label{sec:intro}

On 22 September 2017, the IceCube neutrino observatory detected 
a track-like neutrino event with energy $E_{\nu}\sim 290$ TeV
(IceCube-170922A)  \citep{IceCube:2018cha}. This neutrino event is spatially and temporally associated with
TXS 0506+056, a blazar at a redshift of $z=0.3365\pm 0.0010$ which was in a
flaring state in the $\gamma$-ray energy range at that very moment \citep{Padovani:2018acg}. Extensive follow-up
observations from radio to TeV energy bands revealed that the blazar
TXS 0506+056 was active during this period and enhanced emissions in all
these energy bands were observed, notably the GeV emission is
found to be at high state as observed by $\it Fermi$-LAT \citep{Keivani:2018rnh}. 
On 23 September, $\sim 4$ hours after the neutrino alert, HESS telescopes \citep{Aharonian:2006pe} observed for 1.3 hours and similarly, $\sim 12$ hours after the IceCube-170922A event, the VERITAS telescopes \citep{Holder:2006gi} had a 1-hour follow-up observation in the direction of TXS 0506+056. Both the telescopes also made additional observations on subsequent night with no success. However, the MAGIC telescopes observed very high energy (VHE) $\gamma$-rays above 100 GeV for the first time from TXS 0506+056 on 28 September \citep{Ahnen:2018mvi}.
%Also for the
%first time very high energy (VHE) gamma-rays above
%100 GeV were observed by MAGIC telescopes.
Earlier studies to observe correlation between high-energy neutrinos
and the blazars suffered from poor angular resolution and absence of simultaneous
observation of flares. In 2016, Kadler et al. reported
a PeV neutrino event from the blazar PKS B1424-418 which was
detected by the IceCube neutrino observatory, but it was a shower event
with average median angular error $16^{\circ}$ \citep{Kadler:2016ygj}. 
 
The direct association between a neutrino event, IceCube-170922A, and a point source, TXS 0506+056 was reported
for the first time in multiwavelength observations with high significance  \citep{IceCube:2018cha,Murase:2018iyl}. Several models, particularly leptonic and lepto-hadronic have attempted to explain the observed correlation \citep{Cerruti:2018tmc, Ahnen:2018mvi,Sahakyan:2018voh,Gao:2018mnu,Xue:2019txw,Keivani:2018rnh}.
Most probably, this 
provides direct evidence that active galactic nuclei (AGN) can
accelerate high energy cosmic rays, and produce neutrinos from the
$p\gamma$ and/or $pp$ interactions.

Blazars are a
subclass of AGN and the dominant extra-galactic population in $\gamma$
rays \citep{Acciari:2010aa}, show rapid variability in the entire
electromagnetic spectrum, and have non thermal spectra \citep{Abdo:2010rw}. 
Their 
jet orientation is close to the observer's line of
sight \citep{Urry:1995mg} and powered by matter accretion into the
super massive black hole at the center.
Based on their optical spectra, blazars are divided into flat spectrum
radio quasars (FSRQs) and BL Lac objects (BL Lacs) \citep{Abdo:2009iq}.
The FSRQs are relatively luminous and show strong optical-UV emission
lines, whereas, BL Lacs are less luminous and show only weak emission lines.
The spectral energy distribution (SED) of these blazars
has a double peak structure in the $\nu-\nu F_{\nu}$ plane \citep{Abdo:2009iq}. The low energy peak corresponds to
the synchrotron radiation from a population of relativistic electrons
in the jet and the high energy
peak believed to be either due to the scattering of the high energy electrons with their
self-produced synchrotron photons in the jet (Self-Synchrotron Compton or SSC) \citep{Maraschi:1992iz,Gao:2012sq} or from external sources, such as, photons from the accretion disk, broad line regions, or the dusty torus (External Compton or EC)
\citep{Dermer:1993cz,Sikora:1994zb,Blazejowski:2000ck}. In general, the leptonic models
are very successful in explaining the multiwavelength emission from blazars
\citep{Fossati:1998zn,Ghisellini:1998it,Tavecchio:2010ja,Boettcher:2013wxa}.
Depending on the position of the synchrotron peak, the BL Lac objects
are further divided into low synchrotron peaked  (LSP), intermediate
synchrotron peaked (ISP) and high synchrotron peaked (HSP) blazars respectively \citep{Abdo:2009iq}. For
LSP, the synchrotron peak has frequency $\nu^{peak}_{syn} < 10^{14}$ Hz, for ISP it is in the range $10^{14}\, Hz < \nu^{peak}_{syn}  < 10^{15}$ Hz and for HSP it satisfies
$\nu^{peak}_{syn} > 10^{15}$ Hz. Similarly there is also a shift of
the SSC peak towards higher energy from LSP to HSP.

In the traditional scenario, FSRQs are believed to be promising
sources of high energy neutrinos as they contain high density
photons in the jet and $p\gamma$  process can be effective \citep{Dermer:2014vaa}. But BL Lacs have relatively low photon density in the UV to soft X-ray region hence the $p\gamma$
process to produce neutrinos is not efficient \citep{Righi:2018hhu,Murase:2018iyl}. If TXS 0506+056 is a BL
Lac object, the association of
$\sim 290$ TeV neutrino with it is non trivial to
interpret. HSP blazars have Compton dominance (CD) $\sim
 0.1$ \citep{Padovani:2019xcv}, however, TXS
 0506-056 has CD $\sim 4.5$, implies this may not be an HSP rather a FSRQ, ISP blazar, or LSP blazar. 
Recently, Padovani et
 al. reclassified this as a masquerading BL Lac, namely
a FSRQ with relatively high synchrotron peak \citep{Padovani:2019xcv}.

In this work our motivation is to use the photohadronic model to explain the
VHE $\gamma$-rays and neutrino fluxes from TXS 0506+056 and PKS B1424-418.

\section{Photohadronic model}

Previously, we have shown that the multi-TeV emission from HSP blazars can be
explained very well with the photohadronic model \citep{Sahu:2019kfd}. 
This model relies on
the standard interpretation of the leptonic model to explain both low and
high energy peaks by synchrotron and synchrotron self Compton (SSC) photons respectively as in the
case of any other AGN and blazars. Thereafter, it is assumed that the
flaring occurs within a compact and confined volume of size $R'_f$
inside the blob of radius $R'_b$, with $R'_f < R'_b$  (where $^{\prime}$ implies the jet
co-moving frame and without $^{\prime}$ is in observer frame).  During the flaring,
the compact internal jet is moving slightly faster than the outer one. However, for simplicity, we take their bulk Lorentz factor $\Gamma_{in}\simeq \Gamma_{ext}\simeq
\Gamma$.
Geometrically this represents a double jet structure, one
compact and smaller cone which is enclosed by a bigger one along the
same axis, the geometry of this model is discussed in Fig. 1 of
ref. \citep{Sahu:2015tua}.
Fermi accelerated protons having a
power-law spectrum $dN/dE_p \propto E^{-\alpha}_p$ \citep{Dermer:1993cz} with the power index 
$\alpha \ge 2$ interact with the
background photons in the inner jet region to produce the $\Delta$-resonance which
subsequently decays to $\gamma$-rays via intermediate neutral pion and
to neutrinos through charged pion \citep{Sahu:2012wv}. 
In most of the cases $\alpha=2$ is considered, and for our calculation we also take this value.
The kinematical condition to produce $\Delta$-resonance is
$E_p\epsilon_\gamma=0.32\, \Gamma\, {\cal D}\,  (1+z)^{-2}\, \mathrm{GeV^2}$,
where $E_p$ and $\epsilon_\gamma$ are the
observed proton and seed photon energies respectively; $\Gamma$, ${\cal
  D}$, and $z$ are the bulk Lorentz factor, Doppler factor, and
redshift respectively.
The observed VHE $\gamma$-ray energy is $E_\gamma=0.1 {\cal D}
\Gamma^{-1} E_p$.
In the flaring region we assume
$n'_{\gamma,f}$ is much higher than
the rest of the blob $n'_{\gamma}$ (non-flaring) i.e.
${n'_{\gamma, f}(\epsilon_\gamma)}\gg
{n'_{\gamma}(\epsilon_\gamma)}$. As the inner jet is buried within the
outer jet, it cannot be observed directly. However, by assuming that the Eddington luminosity is equally shared by the jet and the counter jet, the photon density in the inner jet can be constrained to be
$n'_{\gamma,f}\ll L_{Edd}/ {(8\pi R'^2_f\epsilon'_{\gamma})}$ \citep{Sahu:2019lwj}.

The
photon density in the outer region can be calculated from the observed
flux from the SED and, using the scaling behavior,  the
$n'_{\gamma,f}$ can be expressed in terms of $n'_{\gamma}$ \citep{Sahu:2015tua}. 
The outer jet is always there and responsible
for the quiescent state of the blazar while the inner jet is transient and responsible for the flaring event.
In a canonical jet scenario the photohadronic process is inefficient in explaining the multi-TeV emission due to low
photon density. To explain the high energy peak, efficient
acceleration of relativistic protons to ultra-high 
energies in the jet outflow is required and at the same time
the jet kinetic power has to exceed the Eddington luminosity by orders
of magnitude \citep{Cao:2014nia}. However, our compact inner jet scenario eliminates this extreme energy requirement.

The interaction of VHE $\gamma$-rays with the extragalactic background
light (EBL) produces electron-positron pairs and depletes the VHE
$\gamma$-ray flux by a factor of $e^{-\tau_{\gamma\gamma}}$, where
$\tau_{\gamma\gamma}$ is the optical depth for the process
$\gamma\gamma\rightarrow e^+e^-$. To account for the attenuation 
of high energy gamma-rays well known EBL models are used \citep{Dominguez:2010bv,Franceschini:2008tp} and the observed VHE flux is expressed as 
\be
F_{\gamma}(E_{\gamma}) = F_{\gamma,int}(E_{\gamma}) e^{-\tau_{\gamma\gamma}},
\ee
where the intrinsic flux is
\be
F_{\gamma,int}(E_{\gamma})=F_0
\left (   \frac{E_{\gamma}}{TeV}
\right )^{-\delta+3},
\ee
where $\delta=\alpha+\beta$ and $F_0$ is the  normalization
constant determined from the observed VHE SED. During the flaring period, the
background seed photon flux behaves as a power-law $\Phi\propto
E^{-\beta}_{\gamma}$ where $0 < \beta \leq 1.0$ \citep{Sahu:2018zpq}.
Recently, the flaring of HSP blazars have been classified into roughly three
categories depending on the value of $\delta$ \citep{Sahu:2019scf}. Low state emission
corresponds to $\delta=3.0$, high state corresponds to
$2.6<\delta<3.0$, and very high state emission takes place when
$2.5\le \delta \le 2.6$.  As the value of proton spectral index
$\alpha\ge 2$ is known, for different emission states the value of
$\beta$ is constrained accordingly.

\section{Results}

We use the photohadronic model to explain the VHE $\gamma$-ray SED and
estimate the neutrino flux from TXS 0506+056. Using the same approach, we fit the $\gamma$-ray
spectrum of PKS B1424-418 and estimate the neutrino flux.

\subsection{VHE $\gamma$-rays from TXS 0506+056}

On 24 September 2017, the MAGIC telescopes observed TXS 0506+056 under non optimal atmospheric conditions and no $\gamma$-ray were detected. Following the alert of enhanced $\gamma$-ray emission by Fermi-LAT, again MAGIC observed for 13 hours starting 28 September 2017 and detected VHE $\gamma$-rays in the energy range $ 75\, \mathrm{GeV} \le
E_{\gamma} \le 366 \, \mathrm{GeV}$ when it was in a flaring state \citep{Ahnen:2018mvi}.
Taking the
jet bulk Lorentz factor $\Gamma=22$, the viewing angle
$\theta_{view}=0.8^{\circ}$, and the Doppler factor ${\cal D} \simeq 40$, 
MAGIC Collaboration explained the emission using inverse Compton
up-scattering of external photons by accelerated electrons and the
photohadronic interaction. Here we use the photohadronic model to
explain the observed VHE $\gamma$-rays.

In the photohadronic scenario, the VHE SED can be explained very well
by taking $\delta=2.9$,
$F_0=6.0\times 10^{-12}\ \mathrm{erg\ cm^{-2}\ s^{-1}}$
(high state)  and $3.0$, $F_0=5.0\times 10^{-12}\ \mathrm{erg\ cm^{-2}\ s^{-1}}$
(low state) with the EBL correction \citep{Franceschini:2008tp}, as shown in Figure 1. 
Using $\Gamma$ and ${\cal D}$ of MAGIC, the observed VHE
spectrum  in the energy range $ 75\, \mathrm{GeV} \le
E_{\gamma} \le 366 \, \mathrm{GeV}$ is produced from the interaction
of Fermi-acceleration protons in the energy range $750\, GeV \le E_p
\leq 3.7\, TeV$ with the seed photons in energy range  $43 \,
MeV \leq \epsilon_{\gamma} \leq 211\, MeV$
which is in the SSC region. 
In the jet comoving frame the $\gamma$-ray energy $E'_\gamma$ and the seed photon energy $\epsilon'_\gamma$ are respectively in the ranges $2.3\ \mathrm{GeV}\le E'_\gamma \le 12.2 \ \mathrm{GeV}$ and $1.4\ \mathrm{MeV}\le \epsilon'_\gamma \le 7.1 \ \mathrm{MeV}$.
Here we use $R'_f \sim 10^{15}$ cm and $R'_b
\sim 10^{16}-10^{17}$ cm \citep{Ahnen:2018mvi,Keivani:2018rnh}.

The 12.2 GeV photon produced in the inner jet can in principle interact with the seed photons and depletes its energy by producing $e^+e^-$ pairs. However, the mean free path $\lambda_{\gamma\gamma}$ for 12.2 GeV photon interacting with $\epsilon'_\gamma \ge$1.4 MeV seed photon is $\lambda_{\gamma\gamma} > R'_f$ if the photon density is $n'_{\gamma,f}\lesssim 10^{10}\ \mathrm{cm^{-3}}$. This density is also consistent with the moderate efficiency of $\Delta$-resonance process \citep{Sahu:2018gik}, hence, attenuation in $\gamma$-rays in the inner jet due to $e^+e^-$ production is negligible.
It is worth mentioning here that, to fit the observed VHE spectrum of TXS 0506+056, it is not necessary to know the detail of the seed photon density, only the value of $\delta$ is enough to fit it. But to know the range of $\epsilon_{\gamma}$, it is necessary to know the value of ${\cal D}$ and $\Gamma$. Due to the adiabatic expansion of the inner jet, the seed photons with density $n'_{\gamma,f}\lesssim 10^{10}\ \mathrm{cm^{-3}}$ will decrease after crossing into the outer region.

%%%%%%%%%%%%%%%%%%%%%%%%%%%%%%%%%%%%%%%%%%%
\begin{figure}%fig1
%\vspace{-0.3cm}
{\centering
%\resizebox*{0.5\textwidth}{0.5\textheight}
\resizebox*{0.8\textwidth}{0.5\textheight}
{\includegraphics{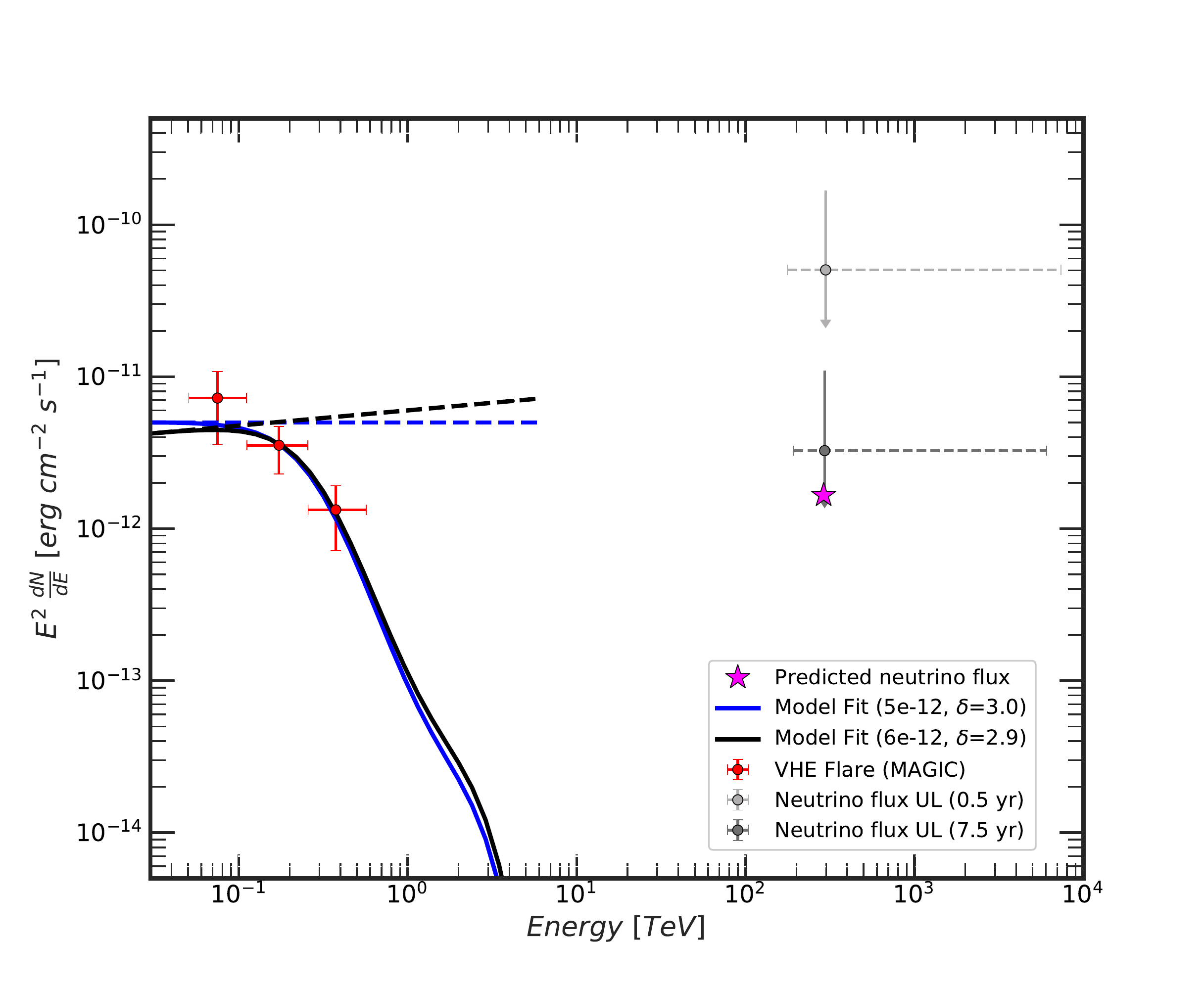}}
\par}
\caption{
The VHE SED of TXS 0506+056 observed by the MAGIC telescopes \citep{Ahnen:2018mvi} starting 28 September 2017 is fitted with the photohadronic model for $\delta=2.9$ and $3.0$ and their respective intrinsic fluxes are shown in dashed curves. 
We have shown the calculated neutrino flux at 290 TeV (magenta star) for $T=158$ days. For comparison, the upper limits (UL) of the neutrino flux for $0.5$ yr and $7.5$ yr estimated by IceCube are also shown \citep{IceCube:2018dnn}.}
\label{fig:figure1}
\end{figure}
%%%%%%%%%%%%%%%%%%%%%%%%%%%%%%%%%%%%%%%%%%%

\subsection{IceCube-170922A neutrino event}

The MAGIC telescopes observed VHE $\gamma$-rays after $\sim$6 days
of the 290 TeV neutrino event \citep{Ahnen:2018mvi, IceCube:2018cha}.
It is possible that during the
neutrino emission period, the flaring was in very high state and in the
next 6 days it slowly decayed to a high or a low state. A similar behavior was
observed from May 1, 2009 flaring of Markarian 501 (Mrk 501) when the flux
increased by a factor of 4 in just 0.5 h (very high state) and afterwards it decreased
but remained in an elevated state for next 2 to 3 days (high
state) \citep{Ahnen:2016hsf}. Had it not been observed during the very high state period, it
would have been assumed that Mrk 501 was in a high state throughout
the above observation period. Keeping this in mind, 
we assume a similar behavior for the flaring of TXS
0506+056. Most probably, on 22 September 2017 the flare was in a very high emission
state with $\delta=2.5-2.6$, when the intrinsic flux might have increased
by order of magnitude in a very short time interval during which the 290 TeV neutrino emission took place through the photohadronic
process. The $\gamma$-ray
energy and its flux subsequently decreased to high state/low 
in next few days, and increase in the
intrinsic flux was mild (high state)/constant (low state) 
as can be seen from Figure 1. 

For $p\gamma\rightarrow \Delta^+$ to take place within the inner
 compact jet region, the time scales should satisfy
\be
t'_{\mathrm{dyn}} < t'_{\mathrm{acc}} < t'_{p\gamma},
\label{timescales}
\ee
where $t'_{\mathrm{dyn}}\simeq R'_f\simeq 3.34\times 10^4 R'_{f,15}\, s$ is
the dynamical time scale, $ t'_{\mathrm{acc}} =10\eta\,E'_p/eB'$ is
acceleration time scale,
$t'_{p\gamma}=(\sigma_{p\gamma}K_{p\gamma} n'_{\gamma,f})^{-1}$
is the $p\gamma$ interaction
time scale, where $K_{p\gamma}=0.2$.
The parameter $\eta$ characterizes the properties of
magnetic disturbances responsible for the acceleration and can vary between
10-100 \citep{Cerruti:2014iwa}.
All other time scales, e.g. $t'_{BH}$ (Bethe-Heitler) and
$t'_{syn}$ (synchrotron) should be larger than $t'_{p\gamma}$.

In the present scenario, 
the Larmour radius of the high energy proton must not exceed the inner jet size
$R'_f\sim 10^{15}$ cm and this corresponds to maximum proton energy energy $E'_{p,max} \simeq
300$ PeV for a magnetic field $B'\sim 1$ G.
The $E_{\nu}=290$ TeV corresponds to the observed
proton energy $E_p\simeq 20 E_{\nu}=5.8$ PeV and in the comoving frame it will be $E'_p=352$ TeV. Correspondingly, the seed photon
energy to produce $\Delta$-resonance will be $\epsilon'_{\gamma} =0.91$ keV in the comoving  frame and in the
observer frame $\epsilon_{\gamma}=27.18$ keV. The seed photons with $\epsilon'_{\gamma}=0.91$ keV and assuming $n'_{\gamma,f}\sim 10^{10}\, \mathrm{cm^{-3}}$ in the inner jet region will expand adiabatically to the outer region of radius $R'_b$ thus decreasing the number density to $n'_{\gamma}\sim 10^{4}\, \mathrm{cm^{-3}}$. The observed flux corresponding to these X-ray photons is estimated to be $F_{keV}\sim 1.4\times 10^{-12}\, \mathrm{erg}\, \mathrm{cm}^{-2}\, \mathrm{s}^{-1}$ and is below the observed limit.

The acceleration time for the proton in the jet is
\be
t'_{acc}=3.9\times 10^4\, \left (
  \frac{\eta}{100} \right ) \left (
  \frac{E'_p}{352\, TeV} \right ) \left (
  \frac{B'}{1\, G} \right )^{-1} \, s.
\ee
The optical depth for the $p\gamma \rightarrow \Delta^+$ process within
the inner jet region is given by
\be
\tau_{p\gamma}=n'_{\gamma,f} R'_{f} \sigma_{p\gamma},
\ee
and we consider $\tau_{p\gamma} \ll 1$, so that excess production
of VHE $\gamma$-rays and neutrinos can be avoided. This corresponds to
$n'_{\gamma,f}\ll 2\times 10^{12}\, \mathrm{cm^{-3}}$. By assuming that the same
neutrino is produced in the outer jet region, the photon density is
estimated to be $n'_{\gamma,f}\simeq  1.5\times 10^{4}\,\mathrm{
cm^{-3}}$. So, the photon density in the jet is constraint to be
\be
1.5\times 10^{4}\,cm^{-3}\ll n'_{\gamma,f}\ll 2\times 10^{12}\, cm^{-3}.
\ee
Here we consider $ 2\times
10^{8}\,\mathrm{cm^{-3}} \lesssim n'_{\gamma,f} \lesssim 2 \times 10^{11}\,\mathrm{cm^{-3}}$ which corresponds to an
optical depth in the range $10^{-4} \lesssim \tau_{p\gamma} \lesssim
10^{-1}$. By taking $n'_{\gamma,f} \simeq
2\times 10^{11}\,\mathrm{cm^{-3}}$, we obtain $t'_{p\gamma} \simeq 1.7\times 10^6\, s$
and for lower density $t'_{p\gamma}$ will be higher, thus the condition given in Eq. (\ref{timescales}) is satisfied. We also estimated the time scale for the Bethe-Heitler (BH) process
in the inner jet and found that $t'_{BH}>t'_{p\gamma}$.

As the photohadronic process and the Bethe-Heitler (BH) pair production process $p\gamma\rightarrow
pe^+e^-$, take place in the same photon background, in principle, the BH process can compete with the photohadronic process \citep{Cerruti:2018tmc}. However, compared to photohadronic process, the BH process has a lower threshold energy and energy loss by the proton to produce leptons pairs is low as the rest mass of the $e^+e^-\sim 1$ MeV is much smaller than the pion mass $m_{\pi}=135$ MeV. Above the pion production threshold, the photohadronic process is dominant over the BH process. Here, the proton energy $E'_p\simeq 352$ TeV which is much above the pion production threshold in the seed photon background, hence, the main energy loss process from the protons is through photopion process \citep{Berezinsky:1988wi, Geddes:1995sd}.

The non-thermalization of electrons by $e\gamma$ interaction implies $n'_{\gamma, f}< 1.5\times 10^9\, \mathrm{cm^{-3}}$.
We calculate the VHE luminosity  $L_{0.07-0.37 \mathrm{TeV}}\sim 2.2\times 10^{45}\, \mathrm{erg\, s^{-1}}$ and by taking $\tau_{p\gamma}\sim 10^{-2}$, the isotropic proton luminosity is $L_p\sim 1.7\times 10^{48}\, \mathrm{erg\, s^{-1}}$. However, $L_p$ can be modified by changing $\tau_{p\gamma}$ and the proton fraction accelerated to VHE energies. In other photohadronic scenarios, the maximum proton luminosity consistent with the SED is estimated as $L_p^{max}\sim 2\times 10^{50}\, \mathrm{erg\, s^{-1}}$ \citep{Keivani:2018rnh}.

The $290$ TeV neutrino energy corresponds to observed $\gamma$-ray energy $E_\gamma\sim 580$ TeV.
These VHE $\gamma$-rays attenuate by interacting with the low energy seed photons  ($\epsilon_\gamma\sim 46$ eV) in the inner and outer region of the jet to produce $e^+e^-$ pairs. Subsequently these lepton pairs will produce electromagnetic cascades of lower energy in the surrounding photon medium and magnetic field.
Furthermore, such high energy photons will be severely attenuated by EBL before reaching the detector. However, the neutrino will escape the jet carrying the information about the parent proton and seed photon spectra.
Although the cascading process of high energy $e^+e^-$ might have initiated simultaneously along with the IceCube neutrino event, the former was not observed. Also, after $\sim$ 6 days of the neutrino event, VHE $\gamma$-rays were observed by MAGIC telescopes. So, even though, cascade emission from secondary leptons were important, it will neither affect the neutrino flux nor the VHE spectrum.

We assumed that the VHE neutrinos are produced during the
very high flaring emission state when the photon flux has increased
dramatically. Then, it is natural to ask, why VHE neutrinos were not observed from the
flaring of Mrk 501 on May 1,  2009 even though, it was in a very high
state ? It is to be noted that, the maximum energy of the proton
depends on the acceleration time scale and the magnetic field in the jet. For Mrk
501, the flare duration was for about 1.5 h (MJD 54952.35-54952.41)
and
$B'\simeq 0.25$ G, which gives $E'_{p,max}\sim 12$ TeV \citep{Ahnen:2016hsf}.
However, to produce $E'_{p,max}\sim 352$ TeV, as in the case of TXS
0506+056, the very high flaring state has to continue for about half a
day in the presence of $B'\sim 1$ G. Thus, the inner jet in Mrk 501 probably had a low
magnetic field and additionally, the very high state did not continue
longer at a stretch to accelerate the protons to sufficiently high
energy even though the active state of the source was much longer.

\subsection{Neutrino Flux estimation}
The number of neutrino events $N_{\nu}$ observed by IceCube at a time period $T$ is given by
\be
N_{\nu} = T \int_{E^*_1}^{E^*_2} \frac{dN}{dE_\nu} A_{eff}(E_\nu)dE_\nu,
\ee
where $E^*_{1,2}=E_{1,2} (1+z)$ and $A_{eff}$ is the effective
area of neutrino in IceCube \citep{Icecubedata}. The neutrino differential flux in
photohadronic model is a power-law
\be
\frac{dN}{dE_\nu}=A_{\nu} \left (\frac{E_\nu}{E_{0}}\right )^{-\delta+1},
\ee
where $A_{\nu}$ is the normalization constant we take $E_{0}=100$ TeV. We assume that the VHE
neutrinos are produced during the very
high energy flaring state of TXS 0506+056 from the $\pi^+$ decay with 
$2.5\le\delta \le 2.6$. This gives
\be
A_\nu=\frac{N_\nu}{T \int_{E^*_1}^{E^*_2} \left
    (\frac{E_\nu}{E_{0}}\right
  )^{-\delta+1}A_{eff}(E_\nu)dE_\nu}.
\label{anu}
\ee
The integral in the denominator can be evaluated numerically for
different values of $\delta$.
The IceCube observed a single muon neutrino event ($N_\nu=1$) of $E_\nu=290$
TeV. For
$\delta=2.5$, $2.6$ and $A_{eff}$ for muon neutrino with the integration
limits 38 TeV to 7 PeV, we obtain
\be
A_\nu \simeq\frac{1}{T}\times \begin{cases} 5.0\times 10^{-10}\, \mathrm{erg^{-1}
  cm^{-2}}, &  {\delta=2.5} \\ 5.9\times 10^{-10}\, \mathrm{erg^{-1}
  cm^{-2}}, &  {\delta=2.6} \end{cases}.
\ee
The multiwavelength observation of TXS 0506+056 suggests that its
most prolonged active period
was about $\sim 0.5\, -1$ year \citep{Ahnen:2018mvi}. The shortest time period when the most
significant excess of $\gamma$-rays were found is the time window
centered at 22 September 2017 with duration 19 days \citep{IceCube:2018cha}. As the number of events are
proportional to the active phase duration, we consider four
time windows for our analysis, namely $T=19, 60, 158$ and $360$ days consistent with
the IC86 runs \citep{IceCube:2018cha}. 

%%%%%%%%%%%%%%%%%%%%%%%%%%%%%%%%
\begin{table}
\centering
\label{table1}
\begin{tabular*}{\columnwidth}{@{\extracolsep{\fill}}lllll@{}}
%\hline
%\multicolumn{1}{@{}l}{Instrument} &Flux state & Period (in MJD unit) \\
\hline

      % \hline \\ [-10 mm]
      T (days) & $\delta$ & $A_{\nu}$ & $F_{\nu}(290\,TeV)$
       & $F^{int}_{\nu}$ \\ [1 mm]
      \hline 
      $19$ & $2.5$ & $3.06\times 10^{-16}$ & $1.34\times 10^{-11}$ & $1.23\times 10^{-10}$ \\
      \, & $2.6$ & $3.60\times 10^{-16}$ & $1.41\times 10^{-11}$ & $1.12\times 10^{-10}$ \\ 
 \hline 
  $60$ & $2.5$ & $1.14\times 10^{-16}$ & $4.23\times 10^{-12}$ & $3.90\times 10^{-11}$ \\ 
      \, & $2.6$ & $1.32\times 10^{-16}$ & $4.48\times 10^{-12}$ & $3.55\times 10^{-11}$ \\ 
 \hline 
  $158$ & $2.5$ & $3.67\times 10^{-17}$ & $1.61\times 10^{-12}$ & $1.48\times 10^{-11}$ \\ 
      \, & $2.6$ & $4.33\times 10^{-17}$ & $1.70\times 10^{-12}$ & $1.35\times 10^{-11}$ \\
 \hline 
  $360$ & $2.5$ & $1.61\times 10^{-18}$ & $7.05\times 10^{-13}$ & $6.49\times 10^{-12}$ \\
      \, & $2.6$ & $1.90\times 10^{-17}$ & $7.47\times 10^{-12}$ & $5.91\times 10^{-12}$ \\  [1 mm]
      \hline
    \end{tabular*}
    \caption{\small The neutrino normalization constant $A_{\nu}$, neutrino flux
      at $E_{\nu}=290$ TeV, $F_{\nu}(290\, TeV)$ and the integrated
      neutrino flux $F^{int}_{\nu}$ are shown for $\delta=2.5$ and
      $2.6$ at different time windows. $A_{\nu}$ is expressed in
      units of $\mathrm{erg^{-1}cm^{-2}}$ and fluxes are given in units of $\mathrm{erg^{-1}cm^{-2}s^{-1}}$.
  }
%  \end{center}
\end{table}
%%%%%%%%%%%%%%%%%%%%%%%%%%%%%%%%

The neutrino flux is given by
\be
F_\nu({E_\nu})=F_{\nu0}  \left ( \frac{E_\nu}{E_{0}} \right )^{-\delta+3},
\ee
where $F_{\nu0}=A_\nu\,E^2_{0}$.
We calculate the integrated neutrino flux $F_{\nu,int}$ for
different time windows with $\delta=2.5,2.6$ as shown in Table 1.
The predicted neutrino fluxes for different time
windows are within the upper limit reported in \citep{IceCube:2018dnn}. We compare our results with the flux predicted by
MAGIC collaboration at 290 TeV \citep{Ahnen:2018mvi} and find that for $T=158$ days
our values are consistent.

%%%%%%%%%%%%%%%%%%%%%%%%%%%%%%%%%%%%%%%%%%%
\begin{figure}%fig2
%\vspace{-0.3cm}
{\centering
%\resizebox*{0.5\textwidth}{0.5\textheight}
\resizebox*{0.8\textwidth}{0.5\textheight}
{\includegraphics{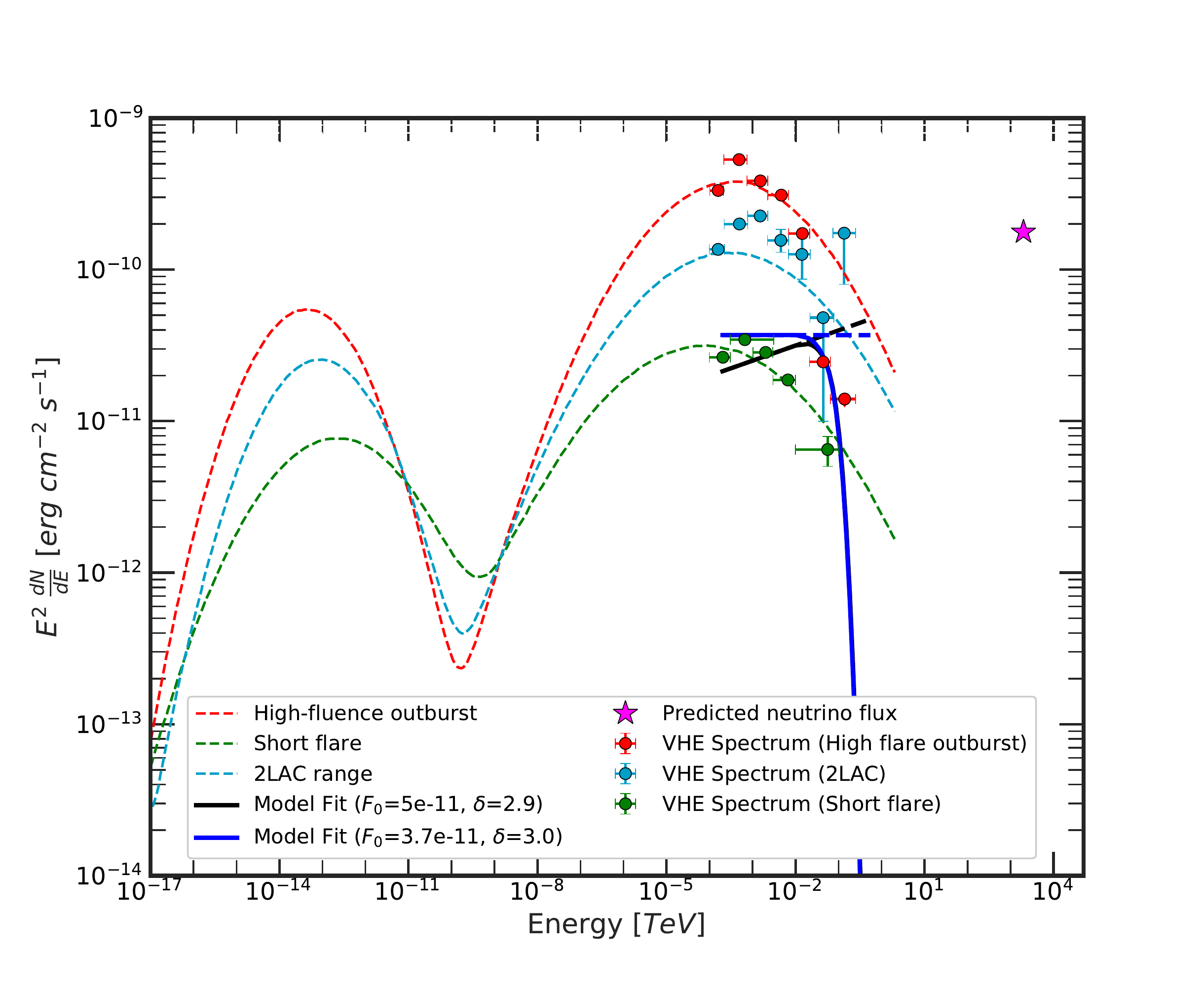}}
\par}
\caption{
The multiwavelength SEDs of PKS B1424-418 observed by different telescopes at different observation periods are shown in Figure 2 of \citep{Kadler:2016ygj}.  The last two points of Fermi-LAT are fitted with the photohadronic model for $\delta=2.9$ and $3.0$ and their respective intrinsic fluxes are shown in dashed curves. The photohadronic prediction of the neutrino flux at 2 PeV for $T=288$ days is shown (magenta star).}
\label{fig:figure2}
\end{figure}
%%%%%%%%%%%%%%%%%%%%%%%%%%%%%%%%%%%%%%%%%%%

\subsection{Neutrino event HESE-35 from PKS B1424-418}

The IceCube has so far detected three shower type
neutrino events in PeV energies, of which
two events are of energy $\sim 1$ PeV and the 
third event (HESE-35) detected on 4 December
2012 at an energy of about 2 PeV \citep{Aartsen:2014gkd}. A spatial and
temporal association of HESE-35 neutrino event with
the flaring FSRQ PKS B1424-418 at a redshift of z=1.522 is suggested
by analyzing the flaring activity in the latter \citep{Kadler:2016ygj}. 
In the time window between 16 July 2012 and 30 April 2013, a period of
$\sim 9$ months, the FSRQ had undergone a major outburst and
$\gamma$-rays in the energy range 100 MeV to 300 GeV were observed
by Fermi-LAT. Also enhanced emission of  X-rays,
optical and radio emissions were observed by different
telescopes \citep{Tavecchio:2013moa}. The arrival time of the 2 PeV
neutrino event coincides with the time window in which the FSRQ had
undergone a major outburst \citep{Kadler:2016ygj}. Using a lepto-hadronic model, with a subdominant
hadronic contribution, the multiwavelength SED is reproduced. It is also shown that the time-wise correlation between the neutrino event and burst phase is weak \citep{Gao:2016uld}.
%, thus suggesting the spatial and temporal correlation.

The SED around the high energy peak (second peak) is due to the SSC
scattering which Fermi-LAT observes. 
During the high-fluence outburst, the spectrum observed by Fermi-LAT
has a sudden change in slope above $\sim 22$ GeV and the last two points do
not fit with two log parabola approximation (Figure 2). It is possible that,
the observed flux above 40 GeV might have different origin than the SSC one, possibly
from neutral
pion decay from the photohadronic process. We fit the VHE flux using
the photohadronic model with $\delta=2.9-3.0$ and  $F_0=(5.0-3.7)\times 10^{-11}\ \mathrm{erg\ cm^{-2}\ s^{-1}}$ in the energy range $43\ \mathrm{GeV}\le E_\gamma\le  139\ \mathrm{GeV}$, as shown in Figure 2. These values of $\delta$
imply that the outburst was either in high emission or in low
emission state.

For PKS B1428-418, we take ${\cal D}=32$, $\Gamma=$20 used by \cite{Tavecchio:2013moa} and consider $R'_b\sim 10^{16}\ \mathrm{cm}$. The non-thermalization condition of electrons by $e\gamma$ interaction implies $n'_{\gamma, f}< 1.5\times 10^8\, \mathrm{cm^{-3}}$.
The VHE luminosity $L_{0.04-0.15 \mathrm{GeV}}\sim 3.1\times 10^{47}\, \mathrm{erg\ s^{-1}}$, which corresponds to an isotropic proton luminosity of $L_p\sim 2.3\times 10^{50} \, \mathrm{erg\, s^{-1}}$ for $\tau_{p\gamma}\sim 10^{-2}$.
Following the same argument as of TXS 0506+056, 
we get that the time scales are
consistent with Eq. (\ref{timescales}) and are given as 
$t'_{\mathrm{dyn}}\simeq 3.3\times 10^{5}\ s$, $t'_{\mathrm{acc}}\simeq 5.6\times 10^{5}\ s$, and $t'_{\mathrm{p\gamma}}\simeq 1.7\times 10^{6}\ s$. To model the SED, \cite{Tavecchio:2013moa} considered a lower magnetic field, however, this modelling does not correspond to the HESE-35 event, thus we consider $B'\sim 1$ G here. The 2 PeV neutrino event ($N_{\nu}=1$) must have originated
from the inner jet of PKS B1424-418 when the flare was in a very high state corresponding to $\delta=2.5-2.6$ and the protons must have accelerated to energy $E_p \simeq 40$ PeV. We calculate the neutrino
flux at $E_{\nu}=2$ PeV and the integrated flux for two time windows $T=288$ days and $T=988$
days consistent with the flaring period of PKS B1424-418 \citep{Kadler:2016ygj}, shown
in Table 2. Our model predicts that, during the major outburst
period $\sim 9$ months the $F_{\nu}(2\, PeV)\sim (1.6-1.9)\times
10^{-10}\,\mathrm{erg\, cm^{-2}\, s^{-1}}$.

%%%%%%%%%%%%%%%%%%%%%%%%%%%%%%%%
\begin{table}
\centering
\label{table2}
\begin{tabular*}{\columnwidth}{@{\extracolsep{\fill}}lllll@{}}
%\hline
%\multicolumn{1}{@{}l}{Instrument} &Flux state & Period (in MJD unit) \\
\hline
 % \hline \\ [-10 mm]
  T (days) & $\delta$ & $A_{\nu}$ & $F_{\nu}(2\,PeV)$ & $F^{int}_{\nu}$ \\ [1 mm]
 \hline 
$288$ & $2.5$ & $1.68\times 10^{-15}$ & $1.93\times 10^{-10}$ & $3.66\times 10^{-10}$ \\
 \, & $2.6$ & $1.87\times 10^{-15}$ & $1.60\times 10^{-10}$ & $3.47\times 10^{-10}$ \\ 
 \hline 
 $988$ & $2.5$ & $4.90\times 10^{-16}$ & $5.62\times 10^{-11}$ & $1.07\times 10^{-10}$ \\ 
 \, & $2.6$ & $5.47\times 10^{-16}$ & $4.65\times 10^{-11}$ & $1.01\times 10^{-10}$ \\
  [1 mm]
  \hline
 \end{tabular*}
 \caption{\small The $A_{\nu}$, $F_{\nu}(2\, PeV)$, and $F^{int}_{\nu}$ are shown for $\delta=2.5$ and
 $2.6$ at different time windows. The units of $A_{\nu}$ and the
 fluxes are the same as given in Table 1.
  }
%  \end{center}
\end{table}
%%%%%%%%%%%%%%%%%%%%%%%%%%%%%%%%

\section{Discussion and Conclusions}

The temporal and directional coincidence of the high energy neutrino
event IceCube-170922A with the flaring blazar TXS 0506+056 in VHE
$\gamma$-rays as well as in low wavelengths suggests that blazars are
strong candidates for at least a fraction of the observed high energy
neutrinos and also VHE cosmic rays and $\gamma$-rays 
\citep{IceCube:2018cha,Kadler:2016ygj}.
To consistently explain this neutrino event and the multiwavelength electromagnetic emission, particularly the VHE $\gamma$-rays observed by MAGIC telescopes, different variant of single-zone leptonic and lepto-hadronic models are used. Here, we briefly discuss some of these models and their results and compare with our model.

\cite{Cerruti:2018tmc} have used the proton synchrotron and SSC emission with a subdominant
but non-negligible contribution from photohadronic cascade to explain the neutrino event. They have shown that the proton–synchrotron picture is disfavored due to insufficient neutrino production rate. On the other hand, to be compatible with the neutrino event, the lepto-hadronic scenario demands more power in the jet. 
Similarly, \cite{Keivani:2018rnh} have proposed a single-zone hybrid lepto-hadronic scenario and shown that $\gamma$-rays are produced by EIC processes and high-energy neutrinos via a radiatively subdominant hadronic component. Here they have argued that, because of the cascade effects, the 0.1–100 keV emissions of TXS 0506+056 are a better probe for the hadronic components than the GeV-TeV emissions.
Yet in another work \cite{Ahnen:2018mvi}, based on the spine-sheath model of \cite{Ghisellini:2004ec}, they have used a single-zone lepto-hadronic scenario where protons and electrons are coaccelerated in the jet and interact with external photons from  the slow moving sheath, to explain the neutrino event and observed VHE $\gamma$-rays. Here it is shown that the VHE $\gamma$-rays are mostly from IC upscattering of external photons by accelerated electrons and the $290$ TeV neutrino event is of photohadronic origin.

In all these above models, apart from many free parameters, it is difficult to explain the VHE $\gamma$-rays and neutrino events in a single-zone scenario, thus multi-zone scenarios may be required. On the other hand, the photohadronic scenario discussed here assumes a composite jet structure with a inner jet of high photon density encircled by an outer jet of lower photon density with similar bulk Lorentz factors ($\Gamma_{in}\simeq\Gamma_{ext}\simeq\Gamma$), and the VHE spectrum can be fitted with a single parameter, the spectral index $\delta$ and the maximum required proton energy is $E_p\simeq 20E_\nu$.

Using photohadronic model, we
have shown that the VHE $\gamma$-ray spectrum observed by MAGIC
telescopes can be explained very well if the flaring was in a high
state. As the 290 TeV neutrino event was observed six days prior to the
gamma-ray event, we argued that, TXS 0506+056 was in a very high
emission state during the neutrino emission period when spectral index
$\delta$ was in the range $2.5-2.6$ and subsequently
decayed to high and low emission states. For the $\Delta$-resonance to
be produced from the $p\gamma$ interaction, we have shown that the
different time scales should satisfy Eq. (\ref{timescales}). As the proton spectral index
is taken to be $\alpha=2$, the power-index $\beta$ of the seed photon in the
inner jet will be in the range $0.5-0.6$. This shows that, the seed
photon flux is flatter in the very high state compared to the one in
the high/low emission state. 
It is the power-law distribution of the seed photon background, having a leptonic origin, decides the nature of the flaring state. 
So there is a direct correlation between the flaring state and the leptonic origin of the seed photons in the jet. As the maximum energy of the proton depends on the acceleration time scale and the magnetic field, in TXS 0506+056 to produce $290$ TeV neutrino, the very high flaring state has to sustain for about half a day in the presence of $B'\sim 1$ G. Similar situation must also prevail for PKS B1424-418 to produce PeV neutrinos.
We took different time windows to estimate the neutrino flux
at 290 TeV and found that our results are consistent with the upper limit reported by IceCube and the estimated flux predicted by MAGIC.
The same method is used to fit the observed VHE $\gamma$-ray
spectrum from PKS B1424-418 and the neutrino flux is estimated.

Although, the IceCube-170922A neutrino event and the flaring of the blazar TXS 0506+056 are found to be correlated, further observation of neutrinos from blazars and follow-up observations in VHE $\gamma$-rays as well as in lower wavelengths are necessary to establish a definitive connection between them. This will also establish AGN as sources of high energy cosmic rays.

The work of S.S. is partially supported by
  DGAPA-UNAM (Mexico) Project No. IN103019. S.N. is partially
  supported by ``JSPS Grants-in-Aid for Scientific Research $<$KAKENHI$>$
  (A) 19H00693'', 
``Pioneering Program of RIKEN for Evolution of Matter in the Universe
(r-EMU)'', and 
``Interdisciplinary Theoretical and Mathematical Sciences Program of RIKEN''.
We are thankful to Alberto Rosales de León, Maxim V. Barkov and Haoning He  for useful discussions. We are also thankful to Anna Franckowiak from IceCube Collaboration for providing us useful information about the neutrino cross section data.
We are thankful to the anonymous referee for his/her insightful comments which helped us to improve the manuscript substantially.

\software{Modelling and plots were done using the distribution of Python 3.7.2. Packages used include: Matplotlib (http://dx.doi.org/10.1109/MCSE.2007.55), Numpy (http://dx.doi.org/10.1109/10.1109/MCSE.2011.37), and Scipy (http://dx.doi.org/10.1109/10.1109/MCSE.2007.58)}. Further calculations were performed using Wolfram Mathematica 12.1 (https://www.wolfram.com/mathematica).

\bibliography{TXS0506_accepted}{}
\bibliographystyle{aasjournal}
\end{document}